\documentclass[12pt]{article}

\usepackage{amsmath}

\newtheorem{theorem}{Theorem}[section]

\newtheorem{definition}{Definition}[section]
\numberwithin{equation}{section}

\begin{document}

\begin{center}
The $\mu$-deformed Segal-Bargmann transform\\
is a Hall type transform
\end{center}

\vskip .5cm

\centerline{Stephen Bruce Sontz\footnote{Research partially
supported by CONACYT (Mexico) project 49187.}}

\centerline{Centro de Investigaci\'on en Matem\'aticas, A.C. (CIMAT)}

\centerline{Guanajuato, Mexico}

\centerline{email: sontz@cimat.mx}

\vskip .8cm

\begin{abstract}
\noindent
We present an explanation of how the $\mu$-deformed Segal-Bargmann spaces, that are
studied in various articles of the author in collaboration with
Angulo, Echavarr\'ia and Pita, can be viewed as deserving their name, that is, how they should be
considered as a part of Segal-Bargmann analysis.
This explanation relates the $\mu$-deformed
Segal-Bargmann transforms to the generalized
Segal-Bargmann transforms introduced by B.~Hall using heat kernel analysis.
All the versions of the $\mu$-deformed Segal-Bargmann transform can be
understood as Hall type transforms.
In particular, we define a $\mu$-deformation of Hall's ``Version~C'' generalized
Segal-Bargmann transform which is then shown to be
a $\mu$-deformed convolution with a $\mu$-deformed heat kernel
followed by analytic continuation.
Our results are generalizations and analogues of the results of Hall.
\end{abstract}
\vskip .3cm
Keywords: Segal-Bargmann analysis, heat kernel analysis, $\mu$-deformed quantum mechanics.
\vskip .3cm \noindent
AMS Subject Classification: primary: 46N50, 47N50,
secondary: 46E15, 81S99
\vskip .3cm

\section{Introduction}
\label{sec1}

We study a deformation of quantum mechanics introduced by Wigner in \cite{WI}.
This deformation depends on a real parameter $\mu  > -1/2$.
Our notation for this parameter is not that used in all the literature, starting
with the original article \cite{WI}.
Our result is about four versions (A through D)
of the $\mu$-deformed Segal-Bargmann transform.
We underline that Version~C of this transform is shown to be a $\mu$-deformed
convolution with a $\mu$-deformed heat kernel followed by analytic continuation.
For those willing to read this result without necessarily being appraised of
all the requisite definitions, we recommend looking at Theorem~\ref{thm2} right away,
which is our main result.
This theorem generalizes and is motivated by the particular case $\mu=0$,
which was presented originally by Hall in \cite{HA1}.

In particular we follow Hall's terminology by referring to various \textit{versions}
of the Segal-Bargmann transform, called Versions A,~B and C in \cite{HA1}.
(We also introduce a Version D, but this is a minor point.)
This terminology however can mislead one into thinking that there is one underlying
object of which various versions (in the usual sense of this word) are being studied.
However, it is true that Versions A and B are closely related,
the only difference between them being a unitary change of variables transform.
This is also the relation between Versions C and D.
See \cite{HA1} and Theorem~\ref{thm2} for exact details about the
various domains, ranges and formulas of these Versions.

Many of our formulas are analogous to formulas in \cite{HA1}.
Thus our first three versions (A to C) of the $\mu$-deformed
Segal-Bargmann transform are Hall type transforms.
Our Version D also fits into this pattern.
This directly relates for the first time the well studied $\mu$-deformed Segal-Bargmann transform
to the seminal work
of Hall, which views Segal-Bargmann analysis as a part of heat kernel analysis.
(See \cite{HA1}, \cite{HA2} and \cite{HA3}.)
In short, we show that $\mu$-deformed Segal-Bargmann analysis
is a part of $\mu$-deformed heat kernel analysis.

   Let us note
there has been much interest and research activity concerning the spaces
and their associated structures that we are studying
here and concerning related, and even more general, spaces and their associated structures.
Besides our work with co-authors (\cite{CAA}, \cite{CAASBS},
\cite{EPS}, \cite{PS}, \cite{PS2}, \cite{PS3} and \cite{SBS})
and the articles by other researchers which will be referenced
later on, we would like to draw attention also to
the relatively recent works on related themes by Sifi and Soltani \cite{SISO},
by Soltani \cite{SO} and by Ben~Sa\"id and {\O}rsted \cite{SBSBO}.
There is also work in progress by Hagedorn \cite{HAG} on a problem in quantum chemistry
where an operator arises that contains $P_\mu^2$, the Dunkl Laplacian, as a term
and so can be thought of as a $\mu$-deformed Hamiltonian.
(See (\ref{Pmudef}) for the definition of the $\mu$-deformed momentum operator $P_\mu$.)
   We also would like to point out that this theory is
connected to probability theory, though this connection will not be used in this article.
As first shown by R\"osler and Voit in \cite{MV},
the Dunkl Laplacian is the generator of a strongly continuous Markov semigroup.
In their article they study this semigroup and its associated stochastic process (a generalization of
Brownian motion, but now with jumps).
More recent work along this line by Gallardo and Yor can be found in \cite{GY1} and \cite{GY2}.

   We start by reviewing some well known results.
See \cite{CAASBS} and \cite{PS2} for a more detailed discussion
of the historical background of this field.
We mention that Marron's thesis \cite{MA} has proven to be quite useful to us.

\begin{definition}
\label{defmumeas}
Let $\lambda > 0$ and $\mu > -1/2$.
Define measures on the complex plane  ${\bf C}$ by
\begin{gather*}
d\nu_{e,\mu,\lambda}(z) := \nu_{e,\mu,\lambda}(z) dx dy, \\
d\nu_{o,\mu,\lambda}(z) := \nu_{o,\mu,\lambda}(z) dx dy,
\end{gather*}
with densities defined for $0 \ne z \in {\bf C}$ by
\begin{gather}
\label{defdene}
\nu_{e,\mu,\lambda}(z) := \lambda 
    \frac{2^{ \frac{1}{2}-\mu }}{\pi \Gamma(\mu+\frac{1}{2} )}
 K_{\mu-\frac{1}{2}} ( | \lambda^\frac{1}{2} z|^2 )
| \lambda^\frac{1}{2} z|^{2\mu +1}  \ , \\
\label{defdeno}
\nu_{o,\mu,\lambda}(z) :=
\lambda \frac{2^{ \frac{1}{2}-\mu }}{\pi \Gamma(\mu+\frac{1}{2} )}
  K_{\mu+\frac{1}{2}} ( | \lambda^\frac{1}{2} z|^2 )
| \lambda^\frac{1}{2} z|^{2\mu +1},
\end{gather}
where $\Gamma$ is the Euler gamma function and $K_\alpha$ is the Macdonald function
of order $\alpha$ as defined in \cite{LEB} as well as in other
standard references such as \cite{AB}.
Lastly, $dxdy$ denotes the Lebesgue measure on ${\bf C}$, the complex plane.
\end{definition}

    The function $K_\alpha$ is also known as the modified
Bessel function of the third kind or Basset's function.
(See \cite{ER}, p.~5.)
But it is also simply known as a modified Bessel function.
(See \cite{GRAD}, p.~961, and \cite{AB}, p. 374.)
An explanation of where the Macdonald functions in Definition \ref{defmumeas}
``come from'' is
given in \cite{SBS}.
The discussion of the Bose-like oscillator in \cite{RO} (especially,
Theorem~5.7) gives an explanation for imposing the condition $\mu > -1/2$,
which we will assume for the rest of this article.

    Let ${\cal H} ({\bf C}) $ be the space of all holomorphic functions
$f : {\bf C} \rightarrow {\bf C}$ of the complex plane to itself.
We note that $f_e := (f + Jf)/2$ (respectively, $f_o := (f - Jf)/2$) defines the even (respectively, odd) part of
$f$, where $Jf(z):= f(-z)$ for all $z\in {\bf C}$ is the parity operator.
So, $ f = f_e + f_o$ and $ Jf = f_e - f_o$ follow.

Throughout the article we use the standard notations for $L^2$ spaces,
for their inner products and for their associated norms.

\begin{definition}
\label{defmusbs}
The \emph{$\mu$-deformed Segal-Bargmann space} for $\lambda > 0$  is
\begin{equation*}
   {\cal B}^2_{\mu,1/\lambda} :=
   {\cal H} ({\bf C})  \cap \left\{ f : {\bf C} \rightarrow
   {\bf C} \ | \  f_e \in L^2( {\bf C}, \nu_{e,\mu,\lambda})
   {\rm ~and~}   f_o \in L^2( {\bf C}, \nu_{o,\mu,\lambda}) \right\},
\end{equation*}
where $ f = f_e + f_o$ is the decomposition of a function into its even and odd parts.
Next we define the norm
$$ 
   || f ||_{ {\cal B}^2_{\mu,1/\lambda} } :=
    \left(  || f_e ||^2_{ L^2( {\bf C}, \nu_{e,\mu,\lambda} )}
            + || f_o ||^2_{ L^2( {\bf C}, \nu_{o,\mu,\lambda} )}
                                                     \right)^{1/2}
$$
for all $ f \in {\cal B}^2_{\mu,1/\lambda}$.
\end{definition}

The reason for using $1/\lambda$ instead of $\lambda$ in the notation
has to do, as we shall see, with maintaining consistency with the notation of Hall in \cite{HA1}.
We have that ${\cal B}^2_{\mu,1/\lambda}$
is a Hilbert space (see \cite{MA}) whose inner product is defined by
\begin{equation} 
\label{defip}
\langle f, g \rangle_{ {\cal B}^2_{\mu,1/\lambda} } 
:= \langle f_e, g_e \rangle_{L^2(\nu_{e,\mu,\lambda})} + \langle f_o, g_o \rangle_{L^2(\nu_{o,\mu,\lambda})}.
\end{equation}

From now on, we write $f = f_e + f_o$ and $g = g_e + g_o$ for the representations of $f$ and $g$
as the sums of their even and odd parts.
When $\mu = 0$ and $\lambda=1$
the space ${\cal B}^2_{\mu,1/\lambda}$ reduces to the usual Segal-Bargmann space, denoted here by $ {\cal B}^2$.
(See \cite{BA, SEG}.)
The motivation for the nomenclature
``Segal-Bargmann space'' in Definition \ref{defmusbs} is given in this article.
Specifically, we shall show that these spaces (and associated structures) conform exactly
to a pattern already identified by Hall in \cite{HA1} in the case $\mu =0$.

   One can relate the parameter $\lambda$ to Planck's constant $\hbar$ by considering
the case $\mu =0$.
We first observe that for $z \in {\bf C}$, $z \ne 0$ and $\mu =0$ we have that
$$
 \nu_{e,0,\lambda}(z) =  \nu_{o,0,\lambda}(z) =
   \lambda \frac{2^{1/2}}{\pi \Gamma(1/2)}
   K_{1/2} ( |\lambda^{1/2} z|^2 ) \cdot |\lambda^{1/2} z| =
   \frac{\lambda}{\pi} e^{- \lambda |z|^2},
$$
which is a normalized Gaussian, using
$K_{1/2}(x) = K_{-1/2}(x) = ( \pi/ (2x) )^{1/2} e^{-x}$
for $x > 0$.
(See \cite{LEB}, p.~110 and p.~112.)
This should be compared with the Gaussian
\begin{equation}
\label{defgauss} 
         \nu_{ {\rm Gauss}, \hbar} (z) := \frac{1}{\pi \hbar} e^{- |z|^2 / \hbar},
\end{equation}
which is the density for the measure of the Segal-Bargmann
space for any $\hbar >0$.
(See \cite{HA2}, p.~9 and p.~21.
Note that the identification $t=\hbar$ is made in \cite{HA2} in the case $\mu=0$.)
So it turns out that $\lambda = 1/\hbar $ for any value of $\mu$ is a reasonable,
though not unique, identification.
Consequently, ${\cal B}^2_{\mu,1/\lambda} = {\cal B}^2_{\mu,\hbar} = {\cal B}^2_{\mu,t}$.
(For those who are confused by the fact that $\hbar$ and $ |z|^2 $
have the same dimensions,
let us note that there is a \emph{normalized} harmonic oscillator Hamiltonian 
implicitly used here.
So both a mass and a frequency have been taken equal to
the dimensionless constant $1$.)

    Note that $\nu_{e,\mu,\lambda}(z) = \lambda \nu_{e,\mu,1}(\lambda^{1/2} z)$ and
$\nu_{o,\mu,\lambda}(z) = \lambda \nu_{o,\mu,1}(\lambda^{1/2} z)$,
so that $\lambda>0$ is a dilation parameter.
Or, in other words, the dilation operator $T_\lambda$ defined by
\begin{equation}
\label{deftlambda}
T_\lambda f (z) := f (\lambda^{1/2} z)
\end{equation}
for $f \in {\cal B}^2_{\mu,1}$ and $ z \in {\bf C}$ 
is a unitary transformation
from ${\cal B}^2_{\mu,1}$ onto ${\cal B}^2_{\mu,1/\lambda}$.
The results of this article hold for every positive value
of the scaling parameter $\lambda$.
However, to keep the notation manageable, we usually will put $\lambda = 1$ hereafter.
Of course, the case of general $\lambda$ is implied by the case $\lambda = 1$
by applying a dilation.
We omit $\lambda$ from the notation when $\lambda=1$.

  While our articles \cite{CAA}, \cite{CAASBS}, \cite{EPS}, \cite{PS}, \cite{PS2}, \cite{PS3} and \cite{SBS}
(most with co-authors) can all be viewed as studies of various properties of a $\mu$-deformation
of standard analysis,
we would like to consider the relation of the structures studied there
with those of standard Segal-Bargmann analysis.
This serves to justify the usage of the terminology ``Segal-Bargmann'' used in those articles.
There are at least three ways (which are not entirely exclusive) for viewing the
theory in the above cited articles
as a mathematical generalization of standard Segal-Bargmann analysis.

    The first way is to be found in the works of Marron \cite{MA}
and Rosenblum \cite{RO,RO2}.
(However, the original idea goes back to a physics paper \cite {WI} by Wigner.)
Their point is that the case $\mu=0$ of their work is precisely the standard
theory originally introduced by Segal in \cite{SEG} and by Bargmann
in \cite{BA} and that every object in standard
Segal-Bargmann analysis has a ``deformation'' for nonzero
$\mu \in (-1/2 , \infty)$.
So they study generalizations, depending on the parameter $\mu$,
of the standard position and momentum operators
of quantum mechanics, denoted as $Q_\mu$ and $P_\mu$.
One way of realizing these operators is
\begin{equation}
\label{Pmudef}
    P_\mu \psi(x) := \frac{\hbar}{i}
    \left(
             \psi^\prime(x) + \frac{\mu}{x} (\psi(x) - \psi(-x) )
    \right) =  \frac{\hbar}{i} D_\mu \psi (x)
\end{equation}
and
\begin{equation}
\label{Qmudef}
     Q_\mu \psi(x) := x \psi(x)
\end{equation}
for suitable functions $\psi : {\bf R} \to {\bf C}$.
(We do not enter into the details of domain considerations here.
We also take $\hbar = 1$ in the rest of this paragraph.)
Up to a multiplicative complex constant the operator $P_\mu$
is the {\em Dunkl operator} $D_\mu$ associated to the Coxeter group
$ {\bf Z}_2 \cong  \{ e, j \}$ of two reflections acting on ${\bf R}$,
where $e$ is the identity map on ${\bf R}$ and $j$ is the reflection in
the origin, namely $j(x) := -x$ for all $x \in {\bf R}$.
See \cite{MR}  and references therein for more on Dunkl operators.
Let us note in passing that this deformation of quantum mechanics is
non-trivial (that is, is not equivalent to the standard case $\mu =0$)
since the canonical commutation relation of quantum mechanics now becomes
\begin{equation}
\label{muccr}
           i [P_\mu, Q_\mu] = I + 2 \mu J,
\end{equation}
where $I$ is the identity operator and $J$ is the parity operator
$J \psi(x) := \psi(-x)$ for $ \psi : {\bf R} \to {\bf C}$ and $x\in {\bf R}$.
While this is not the parity operator introduced above, we abuse notation
by using the same symbol for it.
Equivalently, in terms of the generalized annihilation and
creation operators $a_\mu:= 2^{-1/2} (Q_\mu + i P_\mu)$
and $a^*_\mu:= 2^{-1/2}  (Q_\mu - i P_\mu)$, we have that
$$
       [a_\mu, a^*_\mu] = I + 2 \mu J.
$$

   This $\mu$-deformation of the canonical commutation relation
is just the main idea behind Wigner's article \cite{WI}.
The two operators $a_\mu$ and $a^*_\mu$ are explicitly given in some representation
in \cite{WI}, and so
$Q_\mu$ and $P_\mu$ are already implicit in \cite{WI}.
See \cite{PS3} and the more recent \cite{EPS} for still another way of distinguishing
the standard theory from the $\mu$-deformed theory of quantum mechanics.
Marron and Rosenblum also define and study generalized Hermite polynomials, that are
associated with a generalized harmonic oscillator, as well as a generalized
Segal-Bargmann transform, whose range is precisely the Hilbert space
${\cal B}^2_\mu$ that we have introduced here.
They produce a variety of formulas which have an overall appearance
similar to those
for the case $\mu=0$ and which, furthermore, reduce to the standard
formulas when one substitutes $\mu =0$.
They call this generalization the {\em theory of the Bose-like oscillator,}
though we prefer to call it the {\em $\mu$-deformation
of quantum mechanics} (here, in dimension one).
See \cite{MA} and \cite{RO} for more details about this point of view.
Also, note the antecedents to their work in \cite{CH}, \cite{MU}, \cite{SMMS} and \cite{WI}.
Beware that the $\mu$-deformation should not be confused with
the $q$-deformation of quantum mechanics.
See \cite{MAX} and references therein for a discussion of the latter.

      But there is a second way of seeing a relation of this work
with standard Segal-Bargmann analysis.
This will be our approach in Section~\ref{sec2}.
This has to do with the work of Hall that views the standard Segal-Bargmann
analysis of \cite{BA} and \cite{SEG} as a
part of heat kernel analysis.
(See \cite{HA1} for the original presentation of this idea
and \cite{HA3} and \cite{HA4} and references therein for more recent work.) 
This circle of ideas depends on the construction of various objects
that lead up to a sort of Segal-Bargmann transform,
which is then proved to be unitary.
First, a Hilbert space of \emph{all} $L^2$ functions is
defined on a configuration space
with respect to a measure, both of which must be defined.
Then an integral kernel transform is constructed, using a
heat kernel on the configuration space,
which means that there must be a Laplacian defined on the configuration space.
This transform, known as a {\em generalized Segal-Bargmann transform}, maps
the functions of the previously mentioned $L^2$ space
on the configuration space to holomorphic functions
on the corresponding phase space, which is the cotangent bundle
of the configuration space (assumed to be a smooth manifold).
Then a measure is constructed on the phase space again using a heat kernel,
which means that there must be a Laplacian defined on the phase space.
Then it is shown that the transform is a unitary map onto the (closed!) 
subspace of all \emph{holomorphic} functions in the full $L^2$ space of the phase space.
This subspace is then a sort of generalization of the Segal-Bargmann space.
Of course, a complex structure 
has to be introduced as well on the phase space in order to be 
able to speak of holomorphic functions with that space as their domain.

    Hall and later other researchers have constructed and studied such theories,
but generally in a context where the heat equation arises from a geometrically defined Laplacian
operator associated to rather particular types of differentiable manifolds, such
as compact Lie groups and their homogeneous spaces.
In this article, we develop an analogous theory for what
is known as the $\mu$-deformed Segal-Bargmann transform,
except that the heat equation on the configuration space now
involves a Laplacian constructed from a Dunkl operator.
This gives a more algebraic flavor to the theory due to
the presence of a Coxeter group.
However, there is still an analytic
flavor to the theory as is seen in the following theorem. See \cite{RO}
for a proof.
\begin{theorem}
\label{thm1}
For each $\mu > -1/2$, the $\mu$-deformed heat equation on the real line ${\bf R}$
$$
\frac{ \partial \psi}{ \partial t} =  \frac{1}{2} D_\mu^2 \psi 
$$
with initial condition $\phi_0$ at $t=0$ is solved by
\begin{equation*}
   \psi(x,t) = e^{t D_\mu^2 /2} \phi_0 (x) =
        \int_{ {\bf R} } dq \, |q|^{2 \mu} \rho_{\mu,t}(x,q) \phi_0(q)
\end{equation*}
for $x \in {\bf R}$ and $t>0$, where the $\mu$-deformed heat kernel
$\rho_{\mu,t} : {\bf R} \times {\bf R} \to {\bf C}$
is given explicitly by
\begin{equation}
   \rho_{\mu,t}(x,q) = (2t)^{ -(\mu+1/2) } (\Gamma(\mu +1/2))^{-1}
   \exp \left( - \frac{1}{2t} ( x^2 + q^2 ) \right) \exp_\mu \left( \frac{xq}{t} \right).
\end{equation}
\end{theorem}
We will describe this result in more detail in Section~\ref{sec2}, including
a definition of $\exp_{\mu}$ and
an explanation of how the measure $|q|^{2\mu} dq$ arises naturally.
The $\mu$-deformed heat kernel $\rho_{\mu,t}$ is a crucial element in our next result
as well as its (unique) analytic continuation in the first variable,
$\rho_{\mu,t} : {\bf C} \times {\bf R} \to {\bf C}$.
Note that we use the same notation for the analytic continuation as for
the function itself and let context determine the correct interpretation.

We now state our result.
\begin{theorem}
\label{thm2}
Suppose $\mu > -1/2$ is given.
Then there are four versions of the $\mu$-deformed Segal-Bargmann transform,
the first three of which reduce to a version, as given by Hall in \cite{HA1},
of the standard Segal-Bargmann transform when one puts $\mu = 0$.
They are given by the following integral kernel transforms, where
$z \in {\bf C} $, $q \in {\bf R} $ and $ \hbar = t >0$.
(Note that we use the same notation for the kernel function as well as for its
associated integral transform.)
\begin{enumerate}

\item (Version A)
The kernel function
$$
      A_{\mu,t} (z,q) := \frac{  \rho_{\mu,t}(z,q) }{  (\rho_{\mu,t}(0,q) )^{1/2}}
$$
defines an integral transform
\begin{equation}
\label{Amutransdef}
(A_{\mu,t}\psi)(z) := \int_{ {\bf R} } dq \, |q|^{2\mu} \, A_{\mu,t}(z,q) \psi(q)
\end{equation}
for all $\psi \in L^2 ( {\bf R}, |q|^{2\mu} dq )$.
This is a unitary onto operator
$$
A_{\mu,t} : L^2 ( {\bf R}, |q|^{2\mu} dq ) \rightarrow {\cal B}^2_{\mu,t}.
$$
For $t=1$ this reduces to the generalized Segal-Bargmann transform in Marron \cite{MA}
and Rosenblum \cite{RO2}.
For $t=1$ and $\mu=0$ this further reduces to the usual
Segal-Bargmann transform.
(See Bargmann \cite{BA}, Segal \cite{SEG} and Hall \cite{HA1}.)

\item (Version B, or ground state of Version A)
The kernel function
$$
     B_{\mu,t}(z,q) = \frac{\rho_{\mu,t}(z,q)}{\rho_{\mu,t}(0,q) }
$$
defines an integral transform
$$
  B_{\mu,t} \psi (z) = \int_{ {\bf R} } d\rho_{\mu,t}(q) \, B_{\mu,t}(z,q) \psi(q)
$$
for all $\psi \in L^2 ( {\bf R}, d\rho_{\mu,t} )$,
where $ d\rho_{\mu,t} (q) :=  \rho_{\mu,t} (0,q) |q|^{2\mu} dq$.
This is a unitary onto operator
$$
B_{\mu,t} : L^2 ( {\bf R}, d\rho_{\mu,t} ) \rightarrow {\cal B}^2_{\mu,t}.
$$

\item (Version C)
The kernel function
$$
      C_{\mu,t} (z,q) = \rho_{\mu,t}(z,q)
$$
defines an integral transform
$$
   C_{\mu,t} \psi(z) := \int_{ {\bf R} } dq \, |q|^{2\mu} \, C_{\mu,t}(z,q) \psi(q) =
   ( \sigma_{\mu,t} *_\mu \psi ) (z)
$$
for all $\psi \in L^2 ( {\bf R}, |q|^{2\mu} dq )$,
where $\sigma_{\mu,t} (q) := \rho_{\mu,t}(0,q)$ for $q \in {\bf R}$
is a one variable $\mu$-deformed
heat kernel and $*_\mu$ is $\mu$-deformed convolution (defined later).
So ``Version C'' is given by $\mu$-deformed convolution with a $\mu$-deformed
heat kernel followed by analytic continuation.
This is a unitary onto operator
$$
C_{\mu,t} : L^2 ( {\bf R}, |q|^{2\mu} dq ) \rightarrow {\cal C}^2_{\mu,t},
$$
where
$$
{\cal C}^2_{\mu,t}:= \{ f \in {\cal H} ({\bf C}) \,\, | \,\, Gf \in {\cal B}_{\mu, t/2} \}.
$$
Here $G$ is defined by $Gf(z) := f(2z)/M(2z)$ with
$$
M(z):=
\frac{\exp ( -z^2/(4t) )}{2^{ \mu + 1/2 } t^{ \mu/2 + 1/4 } (\Gamma(\mu + 1/2 ) )^{1/2}}
$$
for $z \in {\bf C}$.
Also the inner product on ${\cal C}^2_{\mu,t}$ is defined by
$$
 \left\langle f_1, f_2 \right\rangle_{ {\cal C}^2_{\mu,t} } :=
\langle Gf_1, Gf_2 \rangle_{ {\cal B}_{\mu,t/2} }
$$
for all $f_1, f_2 \in {\cal C}^2_{\mu,t}$.

   Moreover, there is a relation expressing Version C in terms of Version A.
It is given by
\begin{gather*}
 C_{\mu,t} (z,q) =
\frac{\exp ( -z^2/(4t) )}{2^{ \mu + 1/2 } t^{ \mu/2 + 1/4 } (\Gamma(\mu + 1/2 ) )^{1/2}}
\, A_{\mu,t/2} (z/2,q) \\
 = M(z)  A_{\mu,t/2} (z/2,q).
\end{gather*}

\item (Version D, or ground state of Version C)
The kernel function
$$
      D_{\mu,t} (z,q) =  \frac{  \rho_{\mu,t}(z,q) }{  (\rho_{\mu,t}(0,q) )^{1/2}}
$$
defines an integral transform
$$
  D_{\mu,t} \psi (z) = \int_{ {\bf R} } d\rho_{\mu,t}(q) \, D_{\mu,t}(z,q) \psi(q)
$$
for all $\psi \in L^2 ( {\bf R}, d\rho_{\mu,t} )$.
This is a unitary onto operator
$$
D_{\mu,t} : L^2 ( {\bf R}, d\rho_{\mu,t} ) \rightarrow {\cal C}^2_{\mu,t}.
$$
\end{enumerate}

\end{theorem}

    Let us note that in reference to ``Version A'' we already knew that
$$
A_{\mu,t} (z,q) = \frac{  \rho_{\mu,t}(z,q) }{  (\rho_{\mu,t}(0,q) )^{1/2}} + o(\mu)
$$
as $\mu \to 0$ (pointwise in $(z,q)$) because of the identity proved by Hall in
\cite{HA1} for
the case $\mu=0$ and continuity in the parameter $\mu$.
The point that we are making here is that the $o(\mu)$ term is {\em identically}
equal to zero for all $\mu > -1/2$.
This is quite remarkable, and we did {\em not} expect it to be so.
Similar comments apply to the remaining three versions of this theorem.
Also let us note that Versions A, B and C of Theorem~\ref{thm2} reduce to
results of Hall in \cite{HA1} when one takes $\mu=0$ and are also analogous
to those same results.

    It is a curious fact that Planck's constant $\hbar = t$ appears
in the definition of each version of the Segal-Bargmann transform
as well as in the definition of each of the codomain spaces, although
it appears only in the definition of the domain spaces for Versions B and D.

   Finally there is a third way of seeing a relation of this work
with standard Segal-Bargmann analysis.
This way is to find relations in the $\mu$-deformed theory that do not
depend on the parameter $\mu$.
In particular, the relations for $\mu \ne 0$ are exactly the same as those
for the standard case $\mu=0$.
This is exactly a leitmotif of Theorem~\ref{thm2}.

\section{Proof of Theorem~\ref{thm2}}
\label{sec2}

   The content of this section is a proof of our main result, Theorem~\ref{thm2}.
Along the way we give some definitions and prove some auxiliary results which have
their own independent interest.

We first enter into a more technical discussion of the ideas
leading up to the construction of the four versions in Theorem~\ref{thm2}
of the Segal-Bargmann transform,
using the appropriate heat kernel as was originally done by Hall in \cite{HA1}
in the standard case $\mu=0$.

    We first must specify a configuration space whose phase space should be ${\bf C}$.
So we take the configuration space to be ${\bf R}$,
since the cotangent bundle of ${\bf R}$
can be identified with ${\bf C}$.
Then the ``Schr\"odinger'' space is taken to be the full $L^2$ space on the 
configuration space with respect to some measure. 
This we will take to be
$$
                L^2_\mu := L^2 ( {\bf R}, |q|^{2\mu} dq )
$$
for $\mu > -1/2$.

   The motivation for introducing the measure $|q|^{2\mu} \, dq $ here is that we want
to make $P_\mu$ a symmetric operator (and so $D_\mu$ an anti-symmetric
operator) on $L^2( {\bf R}, \nu(q) dq )$
for some unknown density function $\nu$, that is,
\begin{equation}
\label{unknown_nu}
   \int_{ {\bf R} } dq \, \nu(q) \left( D_\mu \psi(q) \right)^* \phi(q) =
 - \int_{ {\bf R} } dq \, \nu(q) \psi(q)^* D_\mu \phi(q).
\end{equation}
First note that a formal integration by parts gives
\begin{gather*}
  \int_{ {\bf R} } dq \, \nu(q) \left( D_\mu \psi(q) \right)^* \phi(q) =
\int_{ {\bf R} } dq \, \nu(q)
\left( \psi^\prime(q) +\frac{\mu}{q}( \psi(q) - \psi(-q) ) \right)^* \phi(q) \\
=  - \int_{ {\bf R} } \! \! dq \, \psi^*(q) \left(\nu(q) \phi(q)\right)^\prime
  + \int_{ {\bf R} } \! \! dq \, \nu(q) \frac{\mu}{q} \psi^*(q) \phi(q)
  - \int_{ {\bf R} } \! \! dq \, \nu(q) \frac{\mu}{q} \psi^*(-q) \phi(q) \\
= - \int_{ {\bf R} } dq \, \nu(q) \psi^*(q) \phi^\prime(q)
- \int_{ {\bf R} } dq \, \nu^\prime(q) \psi^*(q) \phi(q)
+ \int_{ {\bf R} } dq \, \nu(q) \frac{\mu}{q} \psi^*(q) \phi(q) \\
  + \int_{ {\bf R} } dq \, \nu(-q) \frac{\mu}{q} \psi^*(q) \phi(-q) \\
= \int_{ {\bf R} } dq \, \nu(q) \psi^*(q)
\left[
  - \phi^\prime(q) - \frac{\nu^\prime(q)}{\nu(q)} \phi(q)
  + \frac{\mu}{q}\left( \phi(q) + \phi(-q) \right)
\right]
\end{gather*}
provided that we assume $\nu(q)$ is an even function.
On the other hand, we have that
\begin{gather*}
- \int_{ {\bf R} } dq \, \nu(q) \psi(q)^* D_\mu \phi(q) =
  \int_{ {\bf R} } dq \, \nu(q) \psi(q)^* 
  \left[
          - \phi^\prime(q) - \frac{\mu}{q} (\phi(q) - \phi(-q) )
  \right].
\end{gather*}
So a sufficient condition for (\ref{unknown_nu}) is that
$$
   - \frac{\nu^\prime(q)}{\nu(q)} \phi(q)
  + \frac{\mu}{q} ( \phi(q) + \phi(-q) )= - \frac{\mu}{q} (\phi(q) - \phi(-q))
$$
for all $ 0 \ne q \in {\bf R}$, which in turn is implied by
$$
        \frac{\nu^\prime(q)}{\nu(q)} = \frac{2\mu}{q}
$$
whose general non-negative even solution is $\nu(q) = c |q|^{2\mu}$ for $c>0$.
Now the particular choice of the constant $c$ is not of much importance in this article,
and so we take $c=1$.
It also turns out that the measure $|q|^{2\mu} dq$ is also invariant
with respect to a $\mu$-deformed translation operator or, in other words,
it is a sort of $\mu$-deformation of Haar measure.
We will comment more on this later on.

   Next the $\mu$-deformed Segal-Bargmann transform
$ A_\mu :  L^2_\mu \rightarrow {\cal B}^2_\mu$ is defined for
$\psi \in L^2_\mu$ and $z \in {\bf C}$ by
$$
           A_\mu \psi(z) := \int_{ {\bf R} } dq \, |q|^{2\mu} A_\mu (z,q) \psi(q),
$$
where
\begin{equation}
\label{muSBkerdef}
  A_\mu (z,q) := 2^{-(\mu/2 + 1/4)} \, ( \Gamma ( \mu + 1/2 ))^{-1/2}
                 \exp ( - \frac {1}{2} z^2 - \frac{1}{4} q^2 ) \exp_\mu (qz)
\end{equation}
for $z \in {\bf C}$ and $q \in {\bf R}$. 
This mostly follows Marron and Rosenblum's original formulation in \cite{MA}
and \cite{RO2} respectively, except that their
$\mu$-deformed Segal-Bargmann transform differs from our $A_\mu$ by a unitary
transformation $  L^2_\mu \rightarrow  L^2_\mu$, which is actually a dilation operator.
The fact that $ A_\mu : L^2_\mu \rightarrow {\cal B}^2_\mu$
is a unitary onto transform appears explicitly in \cite{MA},
but was probably already known to Rosenblum.
In our formulation, $A_\mu$ for $\mu=0$ recovers exactly Hall's formulation
for the one-dimensional case ($n=1$) in \cite{HA1}.
In the rest of this article we follow Hall's conventions,
while before we have used Marron and Rosenblum's conventions.

  We now describe in detail what Theorem~\ref{thm1} of Rosenblum
(see \cite{RO}) says.
We warn the diligent reader that we have changed Rosenblum's
formulation to include a factor of $1/2$ in the $\mu$-deformed heat equation.
This is equivalent, of course, to a re-scaling of the ``time'' parameter $t$.
Specifically Rosenblum identified the solution in $ L^2_\mu $
of the $\mu$-deformed heat equation
\begin{gather}
\label{muheateq}
          \frac{ \partial \psi}{ \partial t} =  \frac{1}{2} D_\mu^2 \psi 
\end{gather}
with arbitrary initial condition $ \phi_0 \in L^2_\mu$, that is,
\begin{gather}
\label{initheat}
       \lim_{t \rightarrow 0^+} || \psi( \cdot , t) - \phi_0 ||_{L^2_\mu} = 0,
\end{gather}
to be
\begin{equation}
\label{psixt}
   \psi(x,t) = e^{t D_\mu^2 /2} \phi_0 (x) =
        \int_{ {\bf R} } dq \, |q|^{2 \mu} \rho_{\mu,t}(x,q) \phi_0(q)
\end{equation}
for $x \in {\bf R}$ and $t>0$.
Here $D_\mu$ is {\em the $\mu$-deformed derivative} or {\em Dunkl operator},
which is defined in a suitable domain of functions $\psi: {\bf R} \rightarrow {\bf C}$ by
\begin{gather}
\label{defd}
D_\mu \psi (x) := \psi^\prime (x) + \frac{\mu}{x} \left( \psi \left( x \right) - \psi
\left( -x \right) \right)
\end{gather}
for all $ x \in {\bf R} \setminus \{ 0 \}$.
As noted earlier this is up to a multiplicative complex constant the same as the
$\mu$-deformed momentum operator.
Of course, equation (\ref{defd}) only makes sense as stated for $x \ne 0$.
We also define
$$
D_\mu \psi (0) := \lim_{x \to 0} D_\mu \psi (x) = (2 \mu +1) \psi^\prime (0)
$$
by an application of l'H\^{o}pital's rule for $\psi \in C^1({\bf R})$.

    The expression $e^{t D^2_\mu/2}$ can be understood as being defined by the functional calculus
of the self-adjoint operator $D^2_\mu$, but here it suffices to consider it to be a formal notation.

    Before discussing the formula for the $\mu$-deformed heat kernel
$\rho_{\mu,t}(x,q)$, we wish to introduce a $\mu$-deformed
exponential function $\exp_\mu$.
One way to motivate this is to look for algebraic eigenfunctions
of the operator $D_\mu$, namely a nonzero solution $\phi : {\bf R} \to {\bf C}$ of
\begin{equation}
\label{dmueigen}
           D_\mu \phi(x) = \lambda \phi(x),
\end{equation}
where $\lambda \in {\bf C}$ is the eigenvalue, for all $x \in {\bf R}$.
Assuming that the eigenfunction $\phi$ is a power series in $x \in {\bf R}$,
it is an easy
exercise to show that $\phi(x) = \exp_\mu (\lambda x)$ is a solution where
the {\em $\mu$-deformed exponential function} is defined by
$$
  \exp_\mu(z) := \sum_{n=0}^\infty \frac{1}{\gamma_\mu(n)} z^n
$$
for $z \in {\bf C}$.
Here the {\em $\mu$-deformed factorial function} $\gamma_\mu(n)$ is defined
recursively by $ \gamma_\mu(0) :=1$ and
$\gamma_\mu(n) := (n + 2 \mu \chi_o(n) ) \gamma_\mu (n-1)$ for $n \ge 1$.
Also $\chi_o(n) := 0$ if $n$ is even and $\chi_o(n) := 1$ if $n$ is odd, namely,
$\chi_o$ is the characteristic function of the odd integers.
Let us emphasize that the eigenvalue equation (\ref{dmueigen})
forces this definition of the $\mu$-deformed factorial function.

  It then easily follows that $\exp_\mu(z)$ is well defined
by this power series for all $z \in {\bf C}$ and is an entire function in ${\bf C}$.
Other simple facts are that $\exp_\mu (0) = 1$ and $|\exp_\mu(ix)| \le 1 $
for $\mu \ge 0$ and for all $x \in {\bf R}$.
Note that the results of this paragraph conform with the idea that
the $\mu$-deformation should consist of formulas similar to those in
the standard case $\mu=0$ and, moreover, they should reduce to the
standard case when $\mu=0$ is substituted into them.

    Given this notation, it is known that
\begin{equation}
\label{heatkerdef}
   \rho_{\mu,t}(x,q) = (2t)^{ -(\mu+1/2) } (\Gamma(\mu +1/2))^{-1}
   \exp \left( - \frac{1}{2t} ( x^2 + q^2 ) \right) \exp_\mu \left( \frac{xq}{t} \right)
\end{equation}
is {\em the $\mu$-deformed heat kernel} for
$x \in {\bf R}$, $q \in {\bf R}$ and $t>0$.
We can now note that $\rho_{\mu,t}(x,\cdot) \in L^2_\mu$ for all
$x \in {\bf R}$ follows from equation (\ref{heatkerdef})
and some basic estimates
and therefore the integral in (\ref{psixt}) converges absolutely.
The proof of (\ref{heatkerdef}) given in \cite{RO} uses a straightforward application of
a {\em $\mu$-deformed Fourier transform}, much as one uses the Fourier
transform in the standard case $\mu=0$.
Since the $\mu$-deformed Fourier transform is not used elsewhere in this
article, we omit this material.
See \cite{RO} for all the details.

    Note that in general the $\mu$-deformed heat kernel does not give us a standard
convolution operator in (\ref{psixt}).
However, there is a $\mu$-deformed convolution,
and this is the appropriate concept here.
To introduce this, we first present the following definition
essentially due to Rosenblum in \cite{RO}.
(Our definition actually differs from Rosenblum's by a sign.)

\begin{definition}
Suppose that $\psi : {\bf R} \to {\bf C}$ is such that
$D^n_\mu \psi(q)$ is well defined for
every $q \in {\bf R}$ and every integer $n \ge 1$.
Let $x \in {\bf R}$ be given.
Then the {\em $\mu$-deformed translation of $\psi$ by $x$}
is defined to be
$$
  {\cal T}_{\mu,x} \psi (q) := {\cal T}_{x} \psi (q) :=
     \sum_{n=0}^\infty \frac{(-1)^n}{\gamma_\mu(n)} x^n D^n_\mu \psi(q)
$$
for every $q \in {\bf R}$ such that the infinite series converges absolutely.
\end{definition}

    Note that when we set $\mu=0$ we have that
${\cal T}_{\mu,x}$ is the standard translation operator by $x$,
namely, $({\cal T}_{0,x}\psi)(q)= \psi(q-x)$ provided that $\psi$ is real analytic.
Formally, we have ${\cal T}_{\mu,x} = \exp_\mu(-xD_\mu)$.
Let us note here that the measure $|q|^{2\mu} dq$ is a $\mu$-deformed translation
invariant measure or a {\em $\mu$-deformed Haar measure}, as mentioned earlier.
This means that
$$
     \int_{\bf R} dq \, |q|^{2\mu} \,
   ({\cal T}_{x} \psi )(q) = \int_{\bf R} dq \, |q|^{2\mu} \, \psi (q)
$$
for all $x \in {\bf R}$.
We can see this formally as follows:
\begin{gather*}
     \int_{\bf R} dq \, |q|^{2\mu} \, ({\cal T}_{x} \psi )(q) =
     \int_{\bf R} dq \, |q|^{2\mu} \, \Big( \exp_\mu(-x D_\mu) \psi \Big) (q) \\
= \int_{\bf R} dq \, |q|^{2\mu} \, \left( \sum_{n=0}^\infty \frac{(-1)^n}{\gamma_\mu (n)}
x^n (D^n_\mu\psi )(q) \right) \cdot 1 \\
= \int_{\bf R} dq \, |q|^{2\mu} \, \psi(q) \left( \sum_{n=0}^\infty
\frac{1}{\gamma_\mu (n)} x^n (D^n_\mu 1 )(q) \right)
= \int_{\bf R} dq \, |q|^{2\mu} \, \psi(q),
\end{gather*}
using the anti-symmetry of $D_\mu$ with respect to the measure $|q|^{2\mu} dq$
and $D^n_\mu 1 \equiv 0 $ for all $n \ge 1$.

Another formal argument shows that $|q|^{2\mu}$ is the only
even non-negative density (up to a multiplicative positive constant) that gives a
$\mu$-deformed Haar measure on ${\bf R}$.
Explicitly, if $ \nu(q)$ is such a density, then we have
for all $x \in {\bf R}$ that
$$
\int_{\bf R} dq \, \nu(q) \sum_{n=1}^\infty \frac{(-1)^n}{\gamma_\mu(n)} x^{n}
(D^n_\mu \psi)(q) =0
$$
which implies, after dividing by $x \ne 0$ and then letting $ x \to 0$, that
$$
\int_{\bf R}  dq \, \nu(q) D_\mu \psi(q) =0.
$$
Then a formal integration by parts together with the assumption that $\nu$ is even
leads to the equation $ \nu^\prime (q) = (2 \mu / q ) \nu(q)$, which
in the discussion following (\ref{unknown_nu}) we have already
seen has $\nu(q) = c |q|^{2\mu}$ as its only non-negative even solution, where $c>0$.
We feel that these formal arguments suffice for our purpose, which is
to present some basic properties that are not used to achieve our goal, namely, the
proof of Theorem \ref{thm2}.

\begin{definition}
Let $ \psi_1,\psi_2 : {\bf R} \to {\bf C}$ be measurable functions.
Then we define the {\em $\mu$-deformed convolution of $\psi_1$ and $\psi_2$} by
$$
 (\psi_1 \ast_\mu \psi_2) (x) := \int_{\bf R} dq \, |q|^{2\mu} \,
   ({\cal T}_{q} \psi_1 )(x) \psi_2(q)
$$
for $x \in {\bf R}$ provided that $({\cal T}_{q} \psi_1 )(x)$ exists
for almost all $q \in {\bf R} $ and the integral converges absolutely.
\end{definition}

   This definition reduces to the usual convolution operation of classical analysis
provided that $\mu=0$.

   While we only will need these pointwise defined operations of $\mu$-deformed
translation and convolution, they can be defined on the scale of the Banach
spaces $L^p ( {\bf R}, |q|^{2 \mu} dq)$ for $1 \le p \le \infty$ provided
that $\mu \ge 0$.
Rosenblum's article \cite{RO} contains a whole section devoted to
the $\mu$-deformed translation (which he calls {\em generalized translation}),
but surprisingly there is no mention there of the associated convolution operator.
A generalization of the $\mu$-deformed convolution has been presented in \cite{TX}.

    The following example is central to our argument.
First we evaluate (\ref{heatkerdef}) at $x=0$ to define a $\mu$-deformed
heat kernel $\sigma_{\mu,t}(q)$, a function of one variable $q \in {\bf R} $ by
$$
\sigma_{\mu,t}(q)  := \rho_{\mu,t}(0,q) = (2t)^{ -(\mu+1/2) } (\Gamma(\mu +1/2))^{-1}
   \exp \left( - \frac{1}{2t} q^2 \right)
$$
for $t > 0$ and $\mu > -1/2$.
Then for all $q,x \in {\bf R} $ we have that
\begin{eqnarray*}
  ({\cal T}_{x}  \sigma_{\mu,t} )(q) &=& (2t)^{ -(\mu+1/2) } (\Gamma(\mu +1/2))^{-1}
  \exp \left( - \frac{1}{2t} (q^2 + x^2 ) \right) \exp_\mu \left( \frac{xq}{t} \right)\\
  &=&  \rho_{\mu,t}(x,q)
\end{eqnarray*}
by using formula (4.2.4) in \cite{RO}.
Again we warn the reader that our $\mu$-deformed translation
differs from Rosenblum's generalized translation operator.
See p.~384 in \cite{RO}.

   This says that $\mu$-deformed translations of the $\mu$-deformed
heat kernel $\sigma_{\mu,t}$, a function of one variable,
give the general $\mu$-deformed heat kernel $\rho_{\mu,t}$,
which is a function of two variables.
We immediately get the following result.

\begin{theorem}
\label{thm21}
The solution  (\ref{psixt}) of the $\mu$-deformed heat equation
(\ref{muheateq}) for $x \in\mathbf{R}$, $t > 0$ and with initial condition (\ref{initheat}) is
\begin{gather*}
 \psi_t (x) :=  \psi(x,t) =
        \int_{ {\bf R} } dq \, |q|^{2 \mu} \rho_{\mu,t}(x,q) \phi_0(q) =\\
        \int_{ {\bf R} } dq \, |q|^{2 \mu} \rho_{\mu,t}(q,x) \phi_0(q) =
    \int_{ {\bf R} } dq \, |q|^{2 \mu} ({\cal T}_{q}  \sigma_{\mu,t} )(x) \phi_0(q) =
     \left( \sigma_{\mu,t} \ast_\mu \phi_0 \right) (x),
\end{gather*}
which says that the solution is the $\mu$-deformed convolution of the (one variable)
$\mu$-deformed heat kernel $\sigma_{\mu,t} $ with the initial condition $\phi_0$,
that is, $\psi_t = \sigma_{\mu,t} \ast_\mu \phi_0 $.
\end{theorem}

   Note that this is exactly the same relation that holds when $\mu =0$,
in which case we find with our normalization of the ``time'' parameter $t$ that
$$
 \rho_{0,t}(x,q) = \frac{1}{ (2 \pi t)^{1/2} } \exp \left( -\frac{1}{2t} (x-q)^2 \right)
  = \sigma_{0,t} (x-q),
$$
which is the usual heat kernel in dimension one.

    Now, trying to find analogues of formulas (3) and (4) in \cite{HA1},
we use (\ref{heatkerdef}) to compute that
\begin{equation}
\label{muanalog}
  \frac{  \rho_{\mu,t}(z,q) }{  (\rho_{\mu,t}(0,q) )^{1/2}} =
\frac{\exp ( - z^2/2t  - q^2/4t ) \exp_\mu \left( qz/t \right)}{(2t)^{\mu/2 + 1/4} (\Gamma(\mu + 1/2))^{1/2}},
\end{equation}
where $z \in {\bf C}$, $q \in {\bf R}$ and $t>0$.
We emphasize again that we are using in the numerator on the left
hand side the analytic continuation of
$\rho_{\mu,t}(x,q)$,  $x \in {\bf R}$ to $\rho_{\mu,t}(z,q)$, $z \in {\bf C}$.
Also, we are using in the denominator on the left hand side the fact that
\begin{equation}
\label{mugs}
\rho_{\mu,t}(0,q) =
(2t)^{-(\mu + 1/2)} ( \Gamma(\mu + 1/2) )^{-1} \exp \left( -\frac{1}{2t} q^2 \right)  > 0
\end{equation}
for all $q \in {\bf R}$  and $t>0$.
Taking $t=1$ in (\ref{muanalog}) and using (\ref{muSBkerdef}) we verify that
\begin{equation}
\label{eqalpha}
      A_\mu (z,q)   =   \frac{  \rho_{\mu,1}(z,q) }{  (\rho_{\mu,1}(0,q) )^{1/2}}
\end{equation}
for $z \in {\bf C}$ and $q \in {\bf R}$, which reduces to
equation~(3) in \cite{HA1} when $\mu=0$.
So the kernel of the $\mu$-deformed Segal-Bargmann transform is given in terms of the
$\mu$-deformed heat kernel in exact analogy with the result of Hall in \cite{HA1}.

Moreover, the kernel function defined by
\begin{equation}
\label{eqbeta}
      A_{\mu,t} (z,q) := \frac{  \rho_{\mu,t}(z,q) }{  (\rho_{\mu,t}(0,q) )^{1/2}},
\end{equation}
a straightforward generalization of (\ref{eqalpha}),
itself defines a transformation
\begin{equation}
\label{eqgamma}
        A_{\mu,t} : L^2_{\mu} \rightarrow {\cal B}^2_{\mu,t}
\end{equation}
as given in (\ref{Amutransdef}).
Here, $t > 0$, $z \in \mathbf{C}$ and $q \in \mathbf{R}$.
This definition reduces to the definition (4) in \cite{HA1} when $\mu = 0$.
Note that
$$
     A_{\mu,t} = T_{1/t} A_\mu V_t,
$$
where $V_t : L^2_{\mu} \rightarrow L^2_{\mu}$
is the unitary map defined by $V_t \psi(q) := t^{\mu/2 +1/4} \psi ( t^{1/2} q)$ and
$T_{1/t} : \mathcal{B}_{\mu}^{2} \to \mathcal{B}_{\mu,t}^{2}$ is defined in (\ref{deftlambda}).
This identifies the codomain of $A_{\mu,t}$ and shows it to be a unitary transform
(since it is the composition of three unitary transforms)
with range equal to ${\cal B}^2_{\mu,t}$.
(Recall that $A_\mu : L^2_\mu \to \mathcal{B}_{\mu}^{2}$ is unitary.)

   Hall in \cite{HA1} defines three versions of the original Segal-Bargmann transform.
He uses these in his generalizations to the context of Lie groups and other
types of manifolds.
He denotes these as $A_t$, $B_t$ and $C_t$ where the parameter $t>0$ is identified by him
not only with the ``time'' parameter of a heat equation, but also with
Planck's constant $\hbar$ viewed as a parameter instead of as a fixed physical constant.
Note that this is consistent with the identification of $\lambda$ with $1/t$
in (\ref{eqgamma}) and with our previous identification of $\lambda$ with $1/\hbar$.
We have now finished the discussion of the $\mu$-deformation of the $A_t$ transform,
which we call $A_{\mu,t}$.
This proves Theorem~\ref{thm2}, part 1 (Version A).

    It remains for us to discuss three more new $\mu$-deformations of the Segal-Bargmann transform.
We will denote them as $B_{\mu,t}$, $C_{\mu,t}$  and $D_{\mu,t}$ since the first two
will correspond when $\mu=0$ to $B_t$ and $C_t$ in Hall's formulation.
The $B_{\mu,t}$ transform is just a convenient reformulation of
the $A_{\mu,t}$ transform.
We construct it next, following the method given by Hall in \cite{HA1} 
for the construction of $B_t$ from $A_t$.

   First, we will define a new measure on the
configuration space ${\bf R}$ by using the $\mu$-deformed heat kernel
to change the original measure on the configuration space.
Specifically, in our case we define a probability measure by
\begin{equation}
\label{gsprobmeas}
     d\rho_{\mu,t} (q) :=  \rho_{\mu,t} (0,q) |q|^{2\mu} dq
\end{equation}
for $q \in {\bf R}$.
This can be quite easily checked to be a probability measure by substituting (\ref{mugs})
into (\ref{gsprobmeas}) and integrating over the real line ${\bf R}$.
Then there is a corresponding unitary onto transformation
$$
              U_{\mu,t} : L^2_\mu \rightarrow L^2 ( {\bf R}, d\rho_{\mu,t})
$$
defined for $ \phi \in L^2_\mu $ and $ q \in {\bf R}$ by
$$
 (U_{\mu,t} \phi) (q) := \phi(q)/ (\rho_{\mu,t}(0,q))^{1/2}.
$$
Recall that $\rho_{\mu,t}(0,q) >0$ so that $U_{\mu,t}$ is well defined.
Sometimes $U_{\mu,t}$ is called a {\em change of measure transformation}.
And sometimes one says that $U_{\mu,t}$ takes us to the {\em ground state representation}.
This latter name comes from the fact that $(\rho_{\mu,t}(0,q))^{1/2}$ is the ground state
of the Hamiltonian of the $\mu$-deformed (normalized) harmonic oscillator, which is given
by $H_\mu = (1/2) ( P_\mu^2 + Q_\mu^2) $.
Note that this Hamiltonian depends on $t = \hbar$, even though this is not
indicated in the notation.

   Next we define the transform that is the $\mu$-deformation of the $B_t$ transform
defined in \cite{HA1} by Hall as
$$
  B_{\mu,t} := A_{\mu,t} U_{\mu,t}^{-1} : L^2 ( {\bf R}, d\rho_{\mu,t} )
\rightarrow {\cal B}^2_{\mu,t},
$$
that is, as a composition of two unitary onto transforms.
It follows that $B_{\mu,t}$ is a unitary onto transform.
We compute for $\psi \in L^2 ( {\bf R}, d\rho_{\mu,t} )$ and $z \in {\bf C}$ that
\begin{eqnarray*}
  B_{\mu,t} \psi (z) &=& \int_{ {\bf R} } dq \, |q|^{2\mu} A_{\mu,t}(z,q)
(U_{\mu,t}^{-1} \psi)(q) \\
&=& \int_{ {\bf R} } dq \, |q|^{2\mu}
\frac{  \rho_{\mu,t}(z,q) }{  (\rho_{\mu,t}(0,q) )^{1/2}}
(\rho_{\mu,t}(0,q) )^{1/2} \psi(q) \\
&=& \int_{ {\bf R} } dq \, |q|^{2\mu} \rho_{\mu,t}(z,q) \psi(q).
\end{eqnarray*}
One can use this formula to think of the kernel function of
$ B_{\mu,t}$ as $\rho_{\mu,t}(z,q)$.
This is how Hall describes the situation in the case $\mu =0$ in \cite{HA1}.
However, we prefer to identify the kernel function a different way.
So we write
\begin{eqnarray*}
  B_{\mu,t} \psi (z) &=& \int_{ {\bf R} } dq \, |q|^{2\mu} \rho_{\mu,t}(0,q)
    \left( \frac{\rho_{\mu,t}(z,q)}{\rho_{\mu,t}(0,q) } \right) \psi(q) \\
&=& \int_{ {\bf R} } d\rho_{\mu,t}(q)
    \left( \frac{\rho_{\mu,t}(z,q)}{\rho_{\mu,t}(0,q) } \right) \psi(q)
\end{eqnarray*}
for $\psi \in L^2 ( {\bf R}, d\rho_{\mu,t} )$.
We use this last expression to identify the kernel function as
$$
     B_{\mu,t}(z,q) = \frac{\rho_{\mu,t}(z,q)}{\rho_{\mu,t}(0,q) }
$$
for all $z \in {\bf C}$ and $q \in {\bf R}$.

    So our method is to write the transform as an integral with respect to the measure
that is used for constructing the $L^2$ space that is the ``natural''
domain of the associated integral operator.
In this case, that measure is $d\rho_{\mu,t}(q)$ and not $|q|^{2\mu} dq$.
Having so written the transform, we then identify the coefficient of the function
being transformed as the kernel.

   Of course the transform $B_{\mu,t}$, being a rewritten form of $A_{\mu,t}$,
also reduces in the case $\mu = 0$ to a well-known object,
which is discussed, for example, in \cite{HA1}.
This finishes the proof of Theorem~\ref{thm2}, part~2 (Version B).

   Continuing to follow Hall's presentation,
we now define the $\mu$-deformation of his $C_t$ transform by
\begin{equation}
\label{cmut}
  C_{\mu,t} \psi(z) := \int_{ {\bf R} } dq \, |q|^{2\mu} \rho_{\mu,t}(z,q) \psi(q) =
     \left( \sigma_{\mu,t} \ast_\mu \psi \right) (z)
\end{equation}
for all $\psi \in L^2_\mu$ and $ z \in {\bf C} $.
(The equality holds for $x \in \mathbf{R}$ by Theorem \ref{thm21}.
Then we use analytic continuation.)
Notice that this is exactly the same formula as we obtained for
$B_{\mu,t}$, except that
now the domain Hilbert space is $L^2_\mu$ instead of $L^2 ( {\bf R}, d\rho_{\mu,t} )$.
As such, this is in complete analogy with the relation between $B_t$ and $C_t$ in \cite{HA1}.
And to see this analogy is one reason why we constructed $B_{\mu,t}$.
Accordingly, given our method of identifying kernel functions of integral transforms,
we say that the kernel of $C_{\mu,t}$ is $C_{\mu,t} (z,q) = \rho_{\mu,t}(z,q)$
for $ z \in {\bf C} $ and $ q \in {\bf R} $.
But this is precisely the analytic continuation with respect to the first argument
of the heat kernel $\rho_{\mu,t}(x,q) $ for $ x, q \in {\bf R} $.
So, in analogy with the treatment in \cite{HA1},
the transform $C_{\mu,t}$ can be described as the evolution for ``time'' $t$ of
an element $f \in L^2_\mu$ (given by $\mu$-deformed convolution of $f$ with
the $\mu$-deformed heat kernel at ``time'' $t$), followed by analytic continuation.

   An important question here is whether $C_{\mu,t}$ is really a new object (as $C_t$ is
in the case in \cite{HA1}
when the configuration space is a compact Lie group) or whether  $C_{\mu,t}$ is simply
another reformulation of $A_{\mu,t}$ (as $C_t$ is in the case
when the configuration space is ${\bf R}^n$ for $\mu = 0$ as is shown in \cite{HA1}).
It turns out that the latter is the case, and the clue to resolving this is provided by formula (A.18) in \cite{HA1}.
We claim that formula has a $\mu$-deformation (to be proved a bit later on), namely
\begin{equation}
\label{claim}
  \rho_{\mu,t} (z,q) =
\frac{\exp ( -z^2/(4t) )}{2^{ \mu + 1/2 } t^{ \mu/2 + 1/4 } (\Gamma(\mu + 1/2 ) )^{1/2}}
\frac{\rho_{\mu,t/2} (z/2,q)}{ \sqrt{\rho_{\mu,t/2} (0,q)} },
\end{equation}
which is equivalent in terms of the kernel functions to
\begin{eqnarray}
 C_{\mu,t} (z,q) &=&
\frac{\exp ( -z^2/(4t) )}{2^{ \mu + 1/2 } t^{ \mu/2 + 1/4 } (\Gamma(\mu + 1/2 ) )^{1/2}}  \, A_{\mu,t/2} (z/2,q) \nonumber \\
\label{ACrel}
&=& M(z)  A_{\mu,t/2} (z/2,q).
\end{eqnarray}

   Notice that when we substitute $\mu = 0$ here into (\ref{ACrel}),
we do recover (A.18) in \cite{HA1}, provided that
we also put $n=1$ into (A.18), as we should!
Notice that this fits well into the program of $\mu$-deformations of Rosenblum in that not
only the objects but also their relations have reasonable $\mu$-deformations.
However, it is quite remarkable that we do have an exact equality in (\ref{claim}).
One really only has the right to expect {\em a priori} that the difference of the
two sides of (\ref{claim}) would have a limit equal to zero as $\mu \to 0$.

   So the $A_{\mu,t}$ transform and the $C_{\mu,t}$ transform are related,
though they are not identical.
Firstly, notice that $C_{\mu,t}$ corresponds to $A_{\mu,t/2}$.
Secondly, the phase space point is $z$ for $C_{\mu,t}$ while it is
$z/2$ for $A_{\mu,t/2}$.
The overall factor in (\ref{ACrel}), namely
$$
M(z) =\frac{\exp ( -z^2/(4t) )}{2^{ \mu + 1/2 }
t^{ \mu/2 + 1/4 } (\Gamma(\mu + 1/2 ) )^{1/2}}
$$
does not depend on $q$ and so factors out of the integral in (\ref{cmut}),
and therefore serves as a sort of normalization factor.
However, it does depend on $t$ and $\mu$ as well as on $z$.

   It remains to show the claim (\ref{claim}).
So we compute as follows:
\begin{gather*}
\frac{\rho_{\mu,t/2} (z/2,q)}{ \sqrt{\rho_{\mu,t/2} (0,q)} } =
\frac{  t^{-(\mu+1/2)} (\Gamma(\mu+1/2))^{-1}  e^{ -((z/2)^2 + q^2) /t }
\exp_\mu \left( \frac{qz/2}{t/2} \right) }
{  (t^{-(\mu+1/2)} \Gamma(\mu+1/2)^{-1} e^{- q^2/t} )^{1/2} } \\
= t^{ -(\mu/2 + 1/4) } (\Gamma(\mu+1/2))^{-1/2}
\exp \left( -\frac{z^2}{4t}  -\frac{q^2}{2t} \right)
\exp_\mu \left( \frac{zq}{t} \right)  \\
= t^{ -(\mu/2 + 1/4) } (\Gamma(\mu+1/2))^{-1/2}
\exp \left( -\frac{1}{2t} (z^2 + q^2) \right)
\exp_\mu \left( \frac{zq}{t} \right)
   \exp \left( \frac{z^2}{4t} \right) \\
= t^{ -(\mu/2 + 1/4) } (\Gamma(\mu+1/2))^{-1/2} \exp \left( \frac{z^2}{4t} \right)
(2t)^{\mu + 1/2} \Gamma(\mu+1/2)
   \rho_{\mu,t}(z,q)
\end{gather*}
\begin{gather*}
= 2^{\mu + 1/2} t^{ \mu/2 + 1/4 } (\Gamma(\mu+1/2))^{1/2}
\exp \left( \frac{z^2}{4t} \right) \rho_{\mu,t}(z,q),
\end{gather*}
using (\ref{heatkerdef}), its analytic extension and simple algebra.
This verifies the claim (\ref{claim}).
Continuing with the transform $C_{\mu,t}$, let us note that
\begin{eqnarray*}
    C_{\mu,t} \psi(z) &=& \int_{ {\bf R} } dq \, |q|^{2\mu} C_{\mu,t} (z,q) \psi(q) \\
&=& \int_{ {\bf R} } dq \, |q|^{2\mu} M(z) A_{\mu,t/2} (z/2,q) \psi(q) \\
&=& M(z) \int_{ {\bf R} } dq \, |q|^{2\mu} A_{\mu,t/2} (z/2,q) \psi(q) \\
&=& M(z) ( A_{\mu,t/2} ) \psi(z/2)
\end{eqnarray*}
for all $\psi \in L^2_\mu$ and $ z \in {\bf C} $.
Equivalently, $C_{\mu,t} \psi(2z) = M(2z) ( A_{\mu,t/2} ) \psi(z) $.
So the range of $C_{\mu,t}$, namely all $f$ such that $f = C_{\mu,t} \psi$
for some $\psi \in L^2_\mu$, are exactly those $f \in {\cal H}({\bf C})$
such that $Gf(z) = f(2z)/M(2z)$ lies in the range of $ A_{\mu,t/2}$, which
we have already identified in part~1 of this Theorem to be ${\cal B}^2_{\mu,t/2}$.
This identifies the range of $C_{\mu,t}$ to be ${\cal C}^2_{\mu,t}$ as claimed.
It is now straightforward, using the inner product that we defined on
${\cal C}^2_{\mu,t}$, to show that $C_{\mu,t}$ is unitary.
So, we have now shown Theorem~\ref{thm2}, part~3 (Version C).

   Finally for Theorem~\ref{thm2}, part 4 (Version D), we simply note that the
proof is quite similar to the proof of part~2 (Version B), only now basing
the argument on the result of part~3 (Version C) together with a change of
measure in the configuration space. QED

\section{Conclusion}
A further research avenue is the generalization of this work to
the context of the Segal-Bargmann space associated to a
Coxeter group acting in dimension $n$ (see \cite{SBSBO} and \cite{SO}).
The author has a preprint \cite{SBS2} on this topic.

\section{Acknowledgments}
I wish to thank Andr\'e Martinez and Carlos Villegas for helpful comments
and suggestions.
I am most grateful to Brian Hall for posing the intriguing question of why this
theory should be considered to be a part of Segal-Bargmann analysis.
I thank Miguel Castillo-Salgado for bringing reference \cite{TX} to my attention.
My thanks also go to George Hagedorn for permitting me to mention his work
in progress \cite{HAG}.
I started this article during an academic visit at the Department of Mathematics
of the University of Virginia.
My thanks go to everyone there, and especially to my host Larry Thomas, who have
extended me the warmest hospitality.

\end{document}